# Measurements of strain and bandgap of coherently epitaxially grown wurtzite InAsP-InP core-shell nanowires

D. J. O. Göransson[1,†], M. T. Borgström[1,†], Y. Q. Huang[2,†], M. E. Messing[1], D. Hessman[1], I. A. Buyanova[2], W. M. Chen[2], H. Q. Xu[1,3,4,*]

[1]*NanoLund and Division of Solid State Physics, Lund University, Box 118, S-22100 Lund, Sweden*

[2]*Department of Physics, Chemistry and Biology, Linköping University, S-581 83 Linköping, Sweden*

[3] *Beijing Key Laboratory of Quantum Devices, Key Laboratory for the Physics and Chemistry of Nanodevices, and Department of Electronics, Peking University, Beijing 100871, China*

[4] *Beijing Academy of Quantum Information Sciences, West Bld. #3, No.10 Xibeiwang East Rd., Haidian District, Beijing 100193, China*

[†]*These authors contribute equally to the work.*
[*]*Email addresses: hongqi.xu@ftf.lth.se and hqxu@pku.edu.cn*

**Abstract**

We report on experimental determination of the strain and bandgap of InAsP in epitaxially grown InAsP-InP core-shell nanowires. The core-shell nanowires are grown via metal-organic vapor phase epitaxy. The as-grown nanowires are characterized by transmission electron microscopy, X-ray diffraction, micro-photoluminescence (µPL) spectroscopy and micro-Raman (µ-Raman) spectroscopy measurements. We observe that the core-shell nanowires are of wurtzite (WZ) crystal phase and are coherently strained, with the core and the shell having the same number of atomic planes in each nanowire. We determine the predominantly uniaxial strains formed in the core-shell nanowires along the nanowire growth axis and demonstrate that the strains can be described using an analytical expression. The bandgap energies in the strained WZ InAsP core materials are extracted from the µPL measurements of individual core-shell nanowires. The coherently strained core-shell nanowires demonstrated in this work offer the potentials for use in constructing novel optoelectronic devices and for development of piezoelectric photovoltaic devices.

**Introduction**

Semiconductor nanowires have been the focus of extensive research in recent years [1-11]. Because of the geometry-enabled versatility in bottom up design and of the realization of axial and radial heterostructures, these nanowires offer great potentials in use for the developments of high-performance nanoelectronic and optoelectronic devices [12-22] and of quantum devices [23-28]. It has been demonstrated that semiconductor materials can be combined to form type-I heterostructured, such as InAs-InP, core-shell nanowires, so that charge carriers are confined in the cores, giving an increased carrier mobility in the field-effect transistors (FETs) built from them [15,16]. Here, the shells have been used as a means to reduce surface scattering, which has been a major obstacle for achieving high-performance nanowire FETs [15, 16]. It has also been demonstrated that III-V *pn* junction nanowires, such as radial *pn* junction GaAs-AlGaAs [12] and GaAs-InGaP [13] nanowires, monolithically grown on Si substrates, can be used to construct ultra-small light-emitting diodes. In the nanowire geometry, two materials of largely different lattice constants can be epitaxially grown together to form a coherently strained crystalline heterostructure, containing the same number of atomic planes in both the core and the shell [14-22,29-36]. Methods for spatially resolved measurements of strain in single nanowires, such as convergent beam electron diffraction [37] and focused synchrotron X-ray beam techniques [38], have recently emerged. Investigations of the critical shell thickness for coherent strain in core-shell nanowires have confirmed that the critical thickness is larger than that of a corresponding layer thickness grown on a planar substrate, as predicted by theoretical models [39,40]. It has been predicted that the strain formed coherently in core-shell nanowires can induce a piezoelectric polarization in, e.g., III-V semiconductor materials as a result of the displacement of the relative positions of the group-III and group-V atomic layers, which has potential use in making novel photovoltaic devices [36-42]. Other uses of the effects of coherent strain in core-shell nanowires include strain induced bandgap tuning [18,30,33,34,43] and increased carrier mobilities [33].

Here we report on epitaxial growth of wurtzite (WZ) InAsP-InP core-shell nanowires in a series of shell thicknesses and measurements of strain and optical properties of the nanowires. These materials are of great interest for high speed infrared optoelectronics operated in the telecommunication wavelength window (1.3-1.5 µm), since the bandgap of InAsP can be tuned by compositions from 0.35 to 1.34 eV, corresponding to optical wavelengths from 3.5 µm to 0.9 µm. The WZ InAsP-InP core-shell nanowires are grown by metal-organic vapor phase epitaxy (MOVPE). The lattice parameters of as-grown nanowires are measured by X-ray diffraction (XRD), from which the strain is extracted. We find that the nanowires are coherently strained along the axial [0001] direction. We also show that the strains in the core-shell nanowires at a given core diameter with different shell thicknesses are well described by an analytical formula. The crystal structure and composition of the nanowires are also analyzed by transmission electron microscopy (TEM), micro-photoluminescence (µPL) spectroscopy and micro-Raman (µ-Raman) spectroscopy measurements. The measured µPL spectra show a strain induced blue shift in the emission energy of InAsP cores. Due to the core diameter of ~42 nm and the lattice mismatch of 1.1% between the core and shell materials, misfit dislocations are suppressed by strain relaxation in the shell. Thereby, our work sets a path towards strain-engineered nanowires for applications in novel, high performance optoelectronics and piezoelectronics.

**Experiment**

The InAsP-InP core-shell nanowires are grown via MOVPE. Substrates for growth are prepared from a sulphur doped InP (111)B wafer by depositing Au aerosol particles with a diameter of ~40 nm at a density of ~0.1 µm$^{-2}$ using the method described in Ref. [44]. H$_2$ is used as carrier gas in a total gas flow, including the source gases, set at 6000 sccm. The growth precursors for the InAsP cores are trimethylindium [(CH$_3$)$_3$In], arsine (AsH$_3$) and phosphine (PH$_3$) with molar fractions of $7.1\times10^{-6}$, $2.2\times10^{-5}$, and $6.2\times10^{-3}$, respectively. In order to enhance the formation of the WZ crystal structure of the core, hydrogen sulphide (H$_2$S) is used with a molar fraction of $2.5\times10^{-6}$ [45,46]. In order to impede tapering of the cores, in situ etching with HBr at a molar fraction of $1.1\times10^{-5}$ was used [47].

The growth process of InAsP-InP core-shell nanowires is performed in a low pressure (100 mBar) MOVPE system following the procedure reported in [46]. First, the substrate covered by catalytic Au particles is annealed in the growth chamber in a phosphine-containing flow at 550 °C for 10 min in order to desorb any surface oxides. The temperature is then lowered to 420 °C. The growth of the cores of the nanowires is initiated with a 15-s InP nucleation step by adding trimethylindium to the flow. Wurtzite InAsP cores are then grown for 14 min by adding arsine, HBr and H$_2$S [45] to the flow. Then core growth is interrupted by switching off the flow of arsine, trimethylindium, HBr and H$_2$S. The InP shells are grown by reintroducing the source of trimethylindium to the flow after increasing the reactor temperature to 550 °C and stabilizing the reactor at the temperature for 1 min. The molar fractions of trimethylindium and phosphine are set to $4.0\times10^{-6}$ and $3.7\times10^{-2}$, respectively, for the shell growth. The process yields a high radial growth rate and a low axial growth rate of shells [46]. The growth process is finalized by switching off trimethylindium and the reactor is then cooled down to room temperature under a phosphine/hydrogen flow. In order to vary the strain in the cores, a series of nanowire samples are grown where the InP shell thickness is varied by changing shell growth time. In order to be able to identify any drift of the InAsP composition along the nanowires and between different growth runs, InAsP nanowire reference samples without any shell are also grown.

The grown nanowires were analyzed in a scanning electron microscope (SEM). The diameters of the nanowires are measured by sampling images of about 20 vertically standing nanowires on the center of each substrate. From each nanowire image, the diameter of the nanowire is extracted by measuring the brightness profile along a line perpendicular to the nanowire at the bottom, center and top of the nanowire. A TEM (JEOL 3000F) equipped with a field emission gun is used to analyze the nanowire crystal structure. The composition of the nanowires is measured by energy-dispersive X-ray spectroscopy (EDX) in the scanning TEM (STEM) mode.

The strain and lattice parameter of the nanowires are determined by use of high resolution XRD in the $2\theta - \omega$ configuration. Here, $2\theta$ and $\omega$ are the angles of the detector and sample crystal plane, respectively, with reference to the incident X-ray beam. During the measurements the X-ray beam is held fixed while $2\theta$ and $\omega$ are co-rotated following the relation $\omega = 2\theta/2 + \omega_0$. All samples are initially aligned by adjusting $2\theta$ and $\omega$ to maximize the intensity of the (111) Bragg reflection from the InP substrate. The offset angle of $\omega_0 = 0.1°$ is then introduced to decrease the detected InP substrate reflection intensity while maintaining the intensity originating from the nanowires [48]. In order to correct for alignment errors, the XRD spectra are shifted with the InP (111) peak as an internal reference point which is set to the value $2\theta = 26.281°$. The lattice plane distances along the [0001] direction for the nanowires are then calculated from the $2\theta$ peak positions by Bragg's law. The strain in the axial direction of the core of the nanowires, $\varepsilon$, is calculated by the equation

$$\varepsilon = \frac{c_{cs} - c_{ref}}{c_{ref}}, \quad (1)$$

where $c_{ref}$ is the mean lattice parameter of the reference samples (core only) and $c_{cs}$ is the lattice parameter of the core-shell nanowires.

µPL measurements are carried out to quantify the effect of strain on the bandgap in the InAsP cores of the core-shell nanowires induced by the InP shells. Here, the excitation light from the 659 nm line of a solid state diode laser was focused to a spot of ~1 µm in diameter that covers a single nanowire. The µPL emissions from a total of 40 core-shell nanowires from 6 different samples and a total of 8 reference core nanowires from 4 reference samples are collected and analyzed in this work.

µ-Raman measurements are also carried out to provide supporting information for strain formed in the core-shell nanowires. For these measurements, the nanowires are mechanically transferred on glass substrates. The µ-Raman measurements are done at room temperature in a confocal microscope with excitation by a solid-state laser with emitting wavelength of 659 nm. The laser spot diameter on the substrates is about 0.9 µm. The Raman signals are collected in a backscattering geometry. The collected Raman signals are dispersed with a 0.8-m monochromator and then recorded by a Si CCD camera. We note that the measurements were done without distinguishing the polarization of the phonon modes. A total of 42 InAsP-InP core-shell nanowires from 6 growth samples and 7 InAsP nanowires from a reference sample are measured and analyzed in this work.

**Results and discussion**

Figure 1 shows SEM images of as-grown InAsP-InP core-shell nanowires and reference InAsP nanowires. As shown in Figures 1a and 1b, the reference InAsP nanowires exhibit a uniform diameter along the length of the nanowires. The detailed measurements show that these InAsP nanowires are 42 ± 5 nm in diameter. No parasitic substrate growth is observable. Typical as-grown InAsP-InP core shell nanowires are displayed in Figures 1c-1f. For clarity, a schematic structure of an MOVPE-grown InAsP-InP core-shell nanowire on a growth substrate is shown in Figure 1g. These core-shell nanowires are grown on different substrates with different shell growth times. The shell growth rate is found to be 0.68 nm/s on average and the diameters of these nanowires are in the range of 50 ± 8 to 100 ± 7 nm. Note that slightly smaller diameters are seen at the bottom parts of the InAsP-InP core-shell nanowires and that InP substrate growth is observable at the bases of these core-shell nanowires. This is due to diffusion of growth species from the low surface energy nanowire WZ side facets to the (111)B substrate surface during the shell growth. Figures 1c-1f also show the formation of large tilted particles at the tops of core-shell nanowires, just below or around the Au seed particles, see also the schematic in Figure 1g. These are InP crystalline particles as we will discuss below.

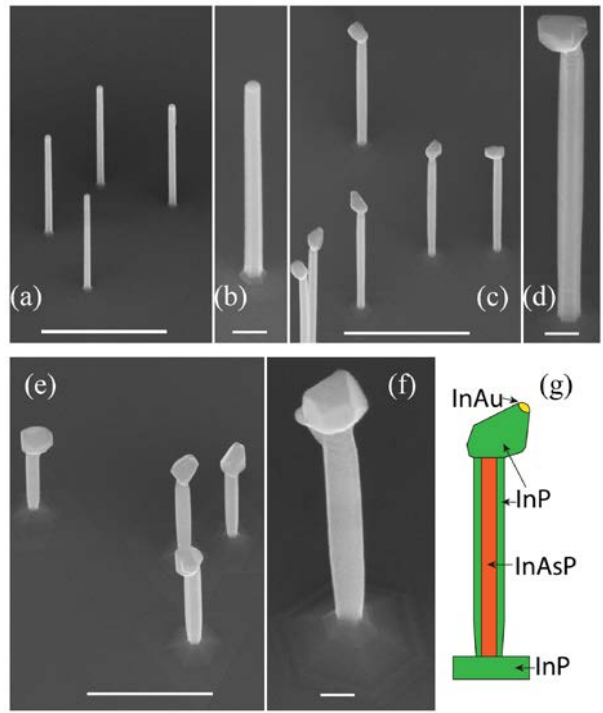

Figure 1. (a) and (b) Low- and high-magnification SEM images of as-grown InAsP nanowires without shells on a growth substrate. (c) and (d) Low- and high-magnification SEM images of as-grown InAsP-InP core-shell nanowires with shells grown in 30 s. (e) and (f) Low- and high-magnification SEM images of as-grown InAsP-InP core-shell nanowires with shells grown in 90 s. (g) Schematic structure of the core-shell nanowires grown for this work. The scale bar in (a), (c) and (e) is 1 µm, while it is 100 nm in (b), (d) and (f). All the images are taken at a tilt angle of 30 ° relative to the surface normal of the growth substrates.

Figure 2 shows the representative results of the structural analysis of our MOVPE-grown InAsP-InP core-shell nanowires. A low-resolution TEM image of an InAsP-InP core-shell nanowire with the shell thickness of 21 nm is shown in Figure 2a, where a tilted particle at the top of the nanowire and a tapered, smaller diameter portion at the bottom of the nanowire are clearly seen. A high angle annular dark field (HAADF) image of the nanowire with EDX scans of As, P and In in the radial direction across the core-shell nanowire is shown in Figure 2b. Here, it is clearly seen that the presence of As is visible only in the core region of the nanowire and any As interdiffusion from the core to the shell is below the detection limit. In order to determine the composition, x in $InAs_xP_{1-x}$ in the cores, a detailed EDX analysis is conducted on two InAsP nanowires from a reference sample without shells with compositional measurements at several locations along the nanowires. The average composition in the InAsP cores is determined to be x = 0.30. A high-resolution TEM image of the nanowire and its corresponding fast Fourier transform are shown in Figure 2c and 2d, from which both the core and shell are found to have WZ crystal structures. Stacking faults along the radial direction across the nanowire is seen to occur with a spacing of 10 to 50 nm in the nanowire. However, these stacking faults do not destroy the coherent lattice connection of the core and shell. The tilted particle on top of the nanowire, below the seed particle, is found to be a pure InP zincblende (ZB) crystal, with a tilted growth direction. The formation of the ZB InP crystalline particle on top of the nanowire could be a result of the thermodynamic processes at the high temperature of 550 °C at which the kinetic growth barriers can be overcome and isotropic growth can be promoted. Note that a small fraction of the nanowires is found to be much longer in length due to rapid axial growth of ZB InP instead of particle formation during shell growth. In the TEM analysis, no formation of dislocations that break the coherent lattice connections of the core and shell is observed and the crystal structure is uniform in the core-to-shell interface as is seen in Figure 2c. However, a dark contrast from moiré patterns is visible in the region where the interface between the core and shell in a nanowire is located. This might arise from varying radial strain in the region, as we will discuss below.

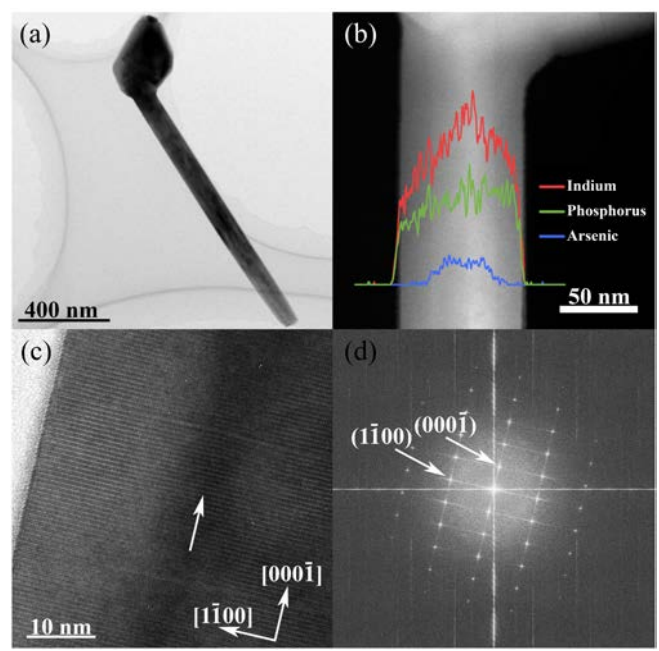

Figure 2. (a) TEM image of an InAsP-InP core-shell nanowire with a shell thickness of 21 nm. The nanowire is imaged along the [$11\bar{2}0$] direction. (b) High angle annular dark field (HAADF) image of the nanowire in (a) and line-scanned EDX spectra across the nanowire. The EDX lines of As, P and In elements are shown in blue, green and red, respectively. (c) and (d) High resolution TEM image of the core-shell nanowire and its corresponding fast Fourier transform. The image in (c) shows that the nanowire has a WZ crystal structure and the white arrow in (c) indicates the moiré pattern formed near the core-shell interface.

Figure 3 shows the XRD spectra of InAsP-InP core-shell nanowires with different shell thicknesses grown on different substrates with different shell growth times and the XRD spectra of pure InAsP nanowires from reference samples. It is seen in Figure 3a that each XRD spectrum consists of two peaks. The right peak originates from the InP substrate and its position should not change with increasing nanowire shell thickness. The other XRD peak is shifted towards higher angle with increasing shell thickness. The shift in the peak position relates to a decrease of the lattice parameter of the core-shell nanowires. It is possible to change the lattice parameter of a core-shell nanowire by straining the core and shell through altering the composition of the InAsP alloy. The reference samples at different growths show only small shifts due to material composition (see Figure 3b) and we conclude that the XRD peak shift in the core-shell nanowire is induced by strain due to lattice mismatched shell growth. Note that the XRD peak profiles show a small tail on the low angle side, suggesting that there is an inhomogeneous strain distribution in the core-shell nanowires. This could be due to the observed varying shell thickness at the bottom part of the nanowires. Furthermore, the peak width does not broaden with increasing shell thickness, indicating that the strain is coherently accommodated in the core-shell nanowires. We also perform XRD measurements on the growth substrates after removal of the nanowires (see Figure S4-1 in the Supporting Information). These measurements confirm that the peaks, which shift toward higher angles, originate from the core-shell nanowires.

From the XRD peak positions of the reference nanowires shown in Figure 3b, the average composition can be calculated by assuming a WZ structure and a linear relation between composition and lattice parameter. By taking the lattice parameters c as 7.0250 Å for WZ InAs [49] and 6.8013 Å for WZ InP [50], the average composition x in the $InAs_xP_{1-x}$ nanowires is found to be x = 0.35. For comparison, we note that the composition values of the InAsP nanowires in one reference sample determined by EDX and XRD are x = 0.30 and x = 0.33, respectively. The discrepancy is within the error of the measurements in EDX and, therefore, we will use the composition values obtained from the XRD data in the strain calculations below. The XRD values correspond to the average composition from each growth substrate, whereas the EDX measurements were performed on a small number of nanowires.

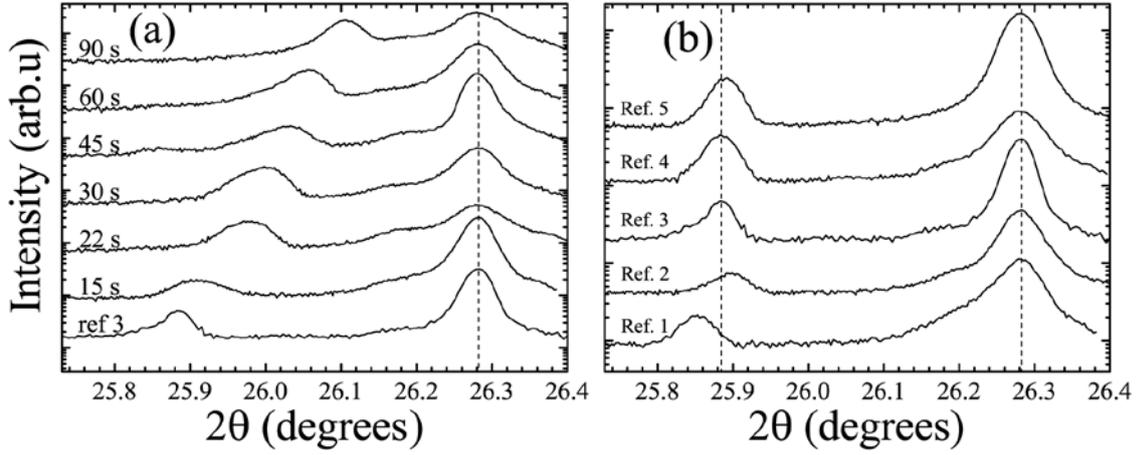

Figure 3. (a) XRD spectra of the core-shell nanowire samples with different shell growth times taken with an $\omega$ offset of -0.1 degrees and using an X-ray wavelength of 1.540593 Å. The peak at 26.281 degrees comes from the ZB InP substrates. The peaks, which shift with increasing shell growth time, from 25.9 to 26.1 degrees, originate from the strained vertically standing nanowires on the substrates. (b) XRD spectra of the InAsP reference nanowires. The dashed vertical lines serve as a guide to the eyes. The spectra are calibrated so that the InP substrate peaks coincide.

The measured strains at different shell thicknesses of the core-shell nanowires are compared to an analytical model [36]. The model was derived for a uniaxial stress along the axis of an infinitely long heterostructure nanowire [36] without taking into consideration the core-shell geometry. The accuracy of the analytical model was checked against numerical 3D finite element method calculations for core-shell nanowires [36,41]. The deviation was found to be about 1%, which confirms the validity of the model. The axial strain ε in the core along the nanowire growth axis is given by

$$\varepsilon = \frac{1 + A_c Y_c / A_s Y_s}{c_c / c_s + A_c Y_c / A_s Y_s} - 1, \qquad (2)$$

where the subscripts $s$ and $c$ refer to shell and core, respectively. $A_c$ and $A_s$ are the cross-section areas, $c_c$ and $c_s$ are the lattice parameters of the materials under unstrained conditions, $Y_c$ and $Y_s$ are the Young's moduli in the [0001] direction. It is seen that the strain depends on the ratios of parameters $c_c/c_s$ and $A_c Y_c / A_s Y_s$. The model is simple and contributes with a direct understanding of how the above parameters affect the axial strain. For the nanowires analyzed here, the area ratio is rapidly decreased when the shell thickness is increased. Figure 4 shows the measured strains in the core-shell nanowires, i.e., the strain values obtained based on the XRD measurements of lattice parameters $c_{cs}$ and $c_{ref}$ according to Eq. (1) and calculated strains based on the analytical model of Eq. (2) for the core-shell nanowires with different shell thicknesses. In the calculations based on the analytical model, the average values of the lattice parameters and composition of the core reference samples measured by XRD was used. The Young's moduli were calculated from the elastic constants of WZ InAs and WZ InP [51]. The elastic constants of InAsP were calculated by the use of Vegard's law. Figure 4 shows that the measured strain as a function of the diameter of the core-shell nanowire follows closely the curve of the analytical model. The discrepancies (within 15%) between the experimental and analytical values are likely to be due to uncertainties of the measurements for the nanowire diameters and compositions of the InAsP cores. The accuracy of the Young's moduli derived from the calculations based on density functional theory can also contribute to the discrepancy.

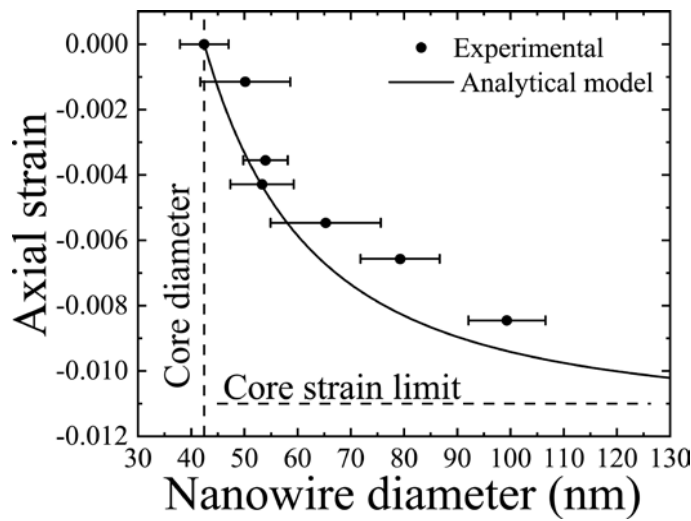

Figure 4. Measured strains along the nanowire axis in the InAsP cores of as-grown InAsP-InP core-shell nanowires plotted against the nanowire diameters. The diameters are measured by SEM and are then calibrated against the diameters of the reference nanowires measured by TEM. The strains are extracted from the XRD measurements of the lattice parameters using Eq. (1). The vertical dashed line marks the diameter of the InAsP nanowires in the reference samples with zero strain. The horizontal dashed line marks core strain limit which would appear in an InAsP-InP core-shell nanowire with an infinitely thick shell. The error bars show the standard deviations of the diameter measurements. The solid line is the theoretical curve obtained using the analytical model of Eq. (2).

Figure 5 shows the extracted peak energies form the µPL spectroscopy measurements of the core-shell nanowires and the reference nanowires. The original measured spectra are shown in Supporting Information as Figures S1-1 to S1-12 for the core-shell samples and as Figures S2-1 to S2-5 for the reference samples. In Figure 5a, the extracted peak energies from the µPL spectroscopy measurements are plotted. Note the blue shift of the PL peak position with increasing optical excitation power, which we attribute to state filling. Therefore, in collecting data for Figure 5a, the excitation power is reduced until the peak position becomes insensitive to the excitation power. We also note that for the reference nanowires with only the InAsP cores, the PL peak energy position exhibits a distribution from 1.055 to 1.097 eV, indicating the existence of composition variations in the InAsP core nanowire samples obtained in different growths. The nanowires with the largest shell thicknesses show an increased PL intensity relative to the reference samples when they were excited with the same power intensity, see the Supporting Information. We attribute this to surface passivation of the optically active cores by the shells, which suppresses carrier recombination via surface states. In addition, due to the relatively large diameter (~40 nm) of the InAsP cores in the core-shell nanowires and the fact that the electron effective mass in the cores is close to that of InP, which give a quantization energy of the order of ~1 meV, no quantization effect is expected to be observed in our µPL spectroscopy measurements.

In Figure 5b, the extracted µPL peak energy is plotted against the evaluated strain in the core-shell nanowires. Here we would like to note that the strain is determined by the XRD measurements which measure ensembles of wires, while the µPL peak energy is extracted from µPL measurements of individual wires where wire-to-wire variations are present. In order to suppress uncertainty arising from wire-to-wire variations, we have made an extensive effort to perform µPL measurements of a sufficiently large number of nanowires from each growth sample and have carried out the comparison of the mean values of the strain from each growth sample determined naturally by XRD to the mean values of the PL energies of many nanowires from the same growth sample, see the results presented in Figure 5a. In this way, fluctuations between NWs on the same growth sample are averaged out to a large extent. In Figure 5b, a trend of increasing µPL energy with increasing strain is seen. We interpret this as an increase in the bandgap of the strained InAsP cores. Here, we should mention that in our µPL measurements, we cannot deduce whether the photon emission is due to band-to-band transition or via exciton decay. But, it should be bandgap related and thus the trend in its energy change with changing shell thickness should reflect correctly the bandgap change in the core with change of strain. By a linear fit to the data points in Figure 5b, a deformation potential of -8.8 eV is extracted for the optically active InAsP cores. This is for the first time that the deformation potential has been experimentally determined for a WZ InAsP structure. Theoretically, the band structures of WZ InAs and InP have been calculated based on first principles methods, and the bandgap and deformation potential of WZ InAsP have been evaluated by linear interpolations between the calculated values for WZ InAs and WZ InP in Ref. [52]. From these results, the theoretical value of change in the bandgap of an InAs$_{0.35}$P$_{0.65}$ NW core with axial strain, $\varepsilon_{zz}$, and in-plane strain, $\varepsilon_{xx} + \varepsilon_{yy}$, can be estimated from $\Delta E_g = -9.47\varepsilon_{zz} - 4.64(\varepsilon_{xx} + \varepsilon_{yy})$ eV. Assuming, a hydrostatic strain in the c-plane, $\varepsilon_{xx} = \varepsilon_{yy}$, and a proportionality factor between the in-plane strain and the strain along the c-axis, $k = (\varepsilon_{xx} + \varepsilon_{yy})/\varepsilon_{zz} = 0.44$, in the core (see Refs. [29, 30, 41, 42] for numerical calculations which justify the assumption), the change in the bandgap of the InAs$_{0.35}$P$_{0.65}$ NW core is found to be $\Delta E_g = -11.5 \times \varepsilon_{zz}$ eV, giving a theoretical deformation potential of -11.5 eV in the core, in reasonably good agreement with our experimental value.

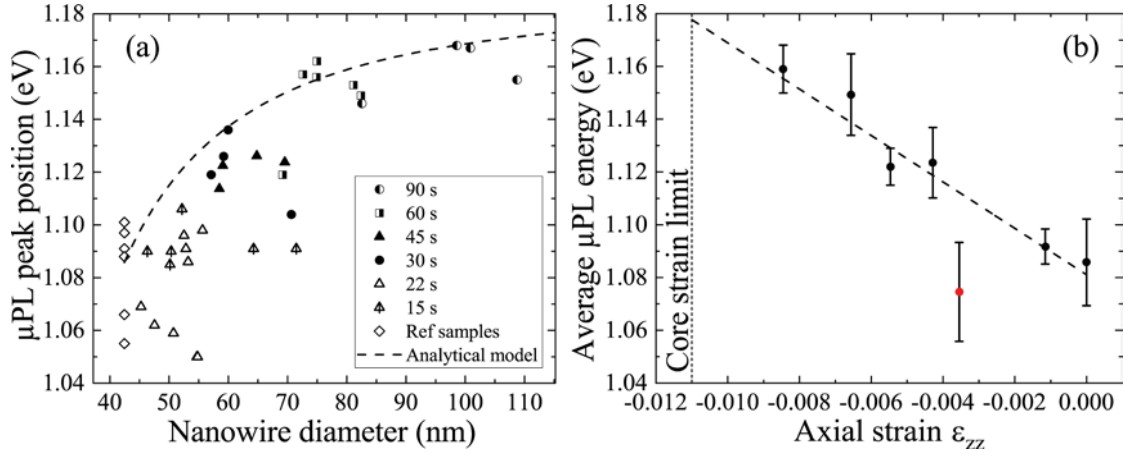

Figure 5. (a) Peak energy positions in the measured µPL spectra of the InAsP-InP core-shell nanowires and the reference InAsP nanowires plotted against the diameters of the nanowires. The individual nanowires measured by the µPL spectroscopy are all measured by SEM to obtain the diameters. The dashed line is a theoretical curve for the expected bandgaps in the strained InAsP cores of the core-shell nanowires with different shell thicknesses with the strains calculated using Eq. (2) by assuming a bandgap deformation potential of -8.8 eV. (b) Averaged µPL peak energy extracted from the measurements shown in (a) as a function of the strain extracted from the XRD measurements for all measured nanowire samples presented in (a). The peak energy of the InAsP reference nanowire samples are shown at zero strain. The core strain limit is shown as a vertical line at an axial strain of 1.1%, corresponding to the case with an infinitely thick shell. The error bars show the standard deviations of the µPL peak positions. The µPL measurements are performed at 4 K under optical excitation at 659 nm. The dashed line corresponds to a linear fit of the experimental data points, excluding the outlier marked by red.

It should be mentioned that the measured PL peak energies of two samples are below the fitted trend line shown in Figure 5b. One is the nanowire sample with shells grown in 45 s and thus an expected strain of -0.55% in the cores. In this sample, the µPL spectra of the nanowires show two bands. We believe that this is because parts of nanowires in the sample are not fully strained and the emission from the unstrained parts gives the lower energy µPL band. The data of this sample presented in Figure 5b are the µPL peak energies originating from the strained part of the nanowires. These are obtained after fitting the two bands by Gaussian peaks and then subtracting the low-energy Gaussian peaks from the measured spectra. The other one is the sample with the nanowire shells grown in 22 s, in which a strain of -0.35% in the cores is expected. We speculate that this low energy emission could be due to local states in the nanowires formed by variations in composition. In the InAsP cores, a 3% variation of the composition can lead to the observed PL energy lowering. A few atomic planes of ZB crystal structure and stacking faults in the nanowires could also be the origin of the low energy emission since the bandgap of ZB InAsP is lower than that in WZ InAsP.

Room temperature µ-Raman spectroscopic measurements are also carried out to provide supporting information for the strains formed in single core-shell NWs (see Section S3 in the Supporting Information for the details of the µ-Raman spectroscopic measurements). As shown in Figure S3-1 of the Supplemental Information, the LO Raman peak is blue shifted with increasing shell growth time, i.e., shell thickness for the NW samples with a small shell thickness. This imply that the compressed strain in the InAsP core is increased with increasing shell thickness, in agreement with the results determined by the XRD measurements. However, for the samples with nanowire diameters in the range of 65-99 nm, the LO peak position is roughly the same. This is different from the XRD measurements but could be understood as follows. First, the interpretation of the LO peak is not trivial in the core-shell nanowires due to the presence of the InP LO peak which is expected to be close to the InAsP LO peak in Raman shift. Second, for the core-shell NW samples with thick NW shells (e.g., in the InP shell thickness range of 18 to 28 nm as in the samples we have measured in this work), we mainly detect the InP shell Raman signals. Thus, as a consequence, although the µ-Raman spectroscopic measurements have provided useful information for the strains formed in the InAsP core regions of InAsP-InP core-shell NWs with thin shells, the core strain determination by the µ-Raman spectroscopic measurements may become unreliable for our NW samples with thick InP shells (see further results and discussion in the Supporting Information).

**Conclusions**

In conclusion, we have studied coherently strained WZ InAsP-InP core-shell nanowires grown by MOVPE. The strains and bandgap energies in the WZ InAsP cores have been determined by the XRD and µPL measurements. We demonstrate that the strains formed in the WZ core-shell nanowires are predominantly uniaxial strains along the nanowire axis and can be well described by an analytical model. The bandgap energies in the strained core InAsP materials are extracted from a large amount of µPL measurements of individual nanowires. In particular, the nanowires with a shell thickness of ~28 nm and a core diameter of ~42 nm show an axial strain of ~0.85 % and a blue shift of the µPL peak of ~80 meV with respect to uncapped nanowires. The core strains in the nanowires are also investigated by µ-Raman measurements. The measurements for our nanowire samples with relatively thinner shells show a trend of blue-shift in LO phonon mode energy with increasing InP shell thickness, in agreement with the XRD measurements. However, we find that it becomes difficult to achieve a reliable core strain determination for our InAsP-InP core-shell nanowires with a thicker InP shell by µ-Raman spectroscopic measurements. Overall, our experimental results, especially the results of the XRD and µPL measurements presented here, demonstrate that the misfit dislocation density is highly suppressed in our InAsP-InP core-shell nanowires and the bandgap of the InAsP core can be effectively tuned by the strain formed in a coherently strained InAsP-InP core-shell nanowire by varying shell thickness. The realization of strained InAsP-InP core-shell nanowires with different diameters, shell thicknesses and compositions is of great interest for optoelectronic, piezoelectronic and photovoltaic applications.

**ASSOCIATED CONTENT**

**Supporting Information**

The Supporting Information is available free of charge on the ACS Publication website.

Further information on the experimental measurements of our MOVPE-grown InAsP-InP core-shell nanowires and reference InAsP nanowires, including the XRD measurements of the lattice structures and the strains in a series of InAsP-InP core-shell nanowire samples, the µPL spectroscopy measurements of individual InAsP-InP core-shell nanowires with different shell thicknesses and reference InAsP nanowires, and the µ-Raman spectroscopy measurements of individual InAsP-InP core-shell nanowires with different shell thicknesses and reference InAsP nanowires. (PDF)

# AUTHOR INFORMATION

**Corresponding author**

*E-mails: hongqi.xu@ftf.lth.se; hqxu@pku.edu.cn

**ORCID**

D. J. O. Göransson: 0000-0001-6963-3330

M. T. Borgström: 0000-0001-8061-0746

M. E. Messing: 0000-0003-1834-236X

H. Q. Xu: 0000-0001-6434-2569

**Notes**

The authors declare no competing financial interest.

# ACKNOWLEDGMENTS

This work was performed within NanoLund with the support from Myfab and was financially supported by the Swedish Research Council (VR), the Ministry of Science and Technology of China through the National Key Research and Development Program of China (Grant Nos. 2017YFA0303304 and 2016YFA0300601), and the National Natural Science Foundation of China (Grant Nos. 11874071, 91221202 and 91421303).

# REFERENCES

[1] Y. Cui & C. M. Lieber, *Science* **291**, 851-853 (2001).

[2] M. H. Huang, S. Mao, H. Feick, H. Yan, Y. Wu, H. Kind, E. Weber, R. Russo & P. Yang, *Science* **292**, 1897-1899 (2001).

[3] Y. Huang, X. Duan, Y. Cui, L. J. Lauhon, K.-H. Kin & C. M. Lieber, *Science* **294**, 1313-1317 (2001).

[4] X. Duan, Y. Huang, Y. Cui, J. Wang & C. M. Lieber, *Nature* **409**, 66-69 (2001).

[5] L. J. Lauhon, M. S. Gudiksen, D. Wang & C. M. Lieber, *Nature* **420**, 57-61 (2002).

[6] Y. Xia, P. Yang, Y. Sun, Y. Wu, B. Mayers, B. Gates, Y. Yin, F. Kim & H. Yan, *Adv. Mater.* **15**, 353-389 (2003).

[7] J. Jiang, W. Lu, Y. Hu, Y. Wu, H. Yan & C. M. Lieber, *Nature* **441**, 489-493 (2006).

[8] B. Tian, X. Zheng, T. J. Kempa, Y. Feng, N. Yu, G. Yu, J. Huang & C. M. Lieber, *Nature* **449**, 885-889 (2007).

[9] H. Yan, H. S. Choe, S.W. Nam, Y. Hu, S. Das, J. F. Klemic, J. C. Ellenbogen & C. M. Lieber, *Nature* **470**, 240-244 (2011).

[10] K. Tomioka, M. Yoshimura & T. Fukui, *Nature* **488**, 189–192 (2012).

[11] J. Wallentin, N. Anttu, D. Asoli, M. Huffman, I. Åberg, M. H. Magnusson, G. Siefer, P. Fuss-Kailuweit, F. Dimroth, B. Witzigmann, H. Q. Xu, L. Samuelson, K. Deppert & M. T. Borgström, *Science* **339**, 1057 (2013).

[12] K. Tomioka, J. Motohisa, S. Hara, K. Hiruma and T. Fukui, *Nano Lett.* **10**, 5, 1639–1644 (2010).

[13] C.P.T. Svensson, T. Mårtensson, J. Trägårdh, C. Larsson, M. Rask, D. Hessman, L. Samuelson & J. Ohlsson, *Nanotechnology* **19**, 30 (2008).

[14] N. Sköld, L. S. Karlsson, M. W. Larsson, M.-E. Pistol, W. Seifert, J. Trägårdh & L. Samuelson, *Nano Lett.* **5**, 10 (2005).

[15] X. Jiang, Q. Xiong, S. Nam, F. Qian, Y. Li & C. M. Lieber, *Nano Lett.* **7**, 3214–3218 (2007).

[16] J. W. W. van Tilburg, R. E. Algra, W. G. G. Immink, M. Verheijen, E. P. A. M. Bakkers & L. P. Kouwenhoven, *Semicond. Sci. Technol.* **25**, 024011 (2010).

[17] O. Salehzadeh, K. L. Kavanagh & S. P. Watkins, *J. Appl. Phys.* **114**, 054301 (2013).

[18] M. Hocevar, L. T. T. Giang, R. Songmuang, M. den Hertog, L. Besombes, J. Bleuse, Y.-M. Niquet & N. T. Pelekanos, *Appl. Phys. Lett.* 102, 191103 (2013).

[19] K. Moratis, S. L. Tan, S. Germanis, C. Katsidis, M. Androulidaki, K. Tsagaraki, Z. Hatzopoulos, F. Donatini, J. Cibert, Y. -M. Niquet, H. Mariette & N. T. Pelekanos, *Nanoscale Res. Lett.* **11**, 176 (2016).

[20] P. Mohan, J. Motohisa & T. Fukui, *Appl. Phys. Lett* **88**, 133105 (2006).

[21] M. H. Hadj Alouane, R. Anufriev, N. Chauvin, H. Khmissi, K. Naji, B. Ilahi, H. Maaref, G. Patriarche, M. Gendry & C. Bru-Chevallier, *Nanotechnology* **22**, 405702 (2011).

[22] J. Treu, M. Bomann, H. Schmeiduch, M. Möblinger, S. Morkötter, S. Matich, P. Wiecha, K. Saller, B. Mayer, M. Bichler, M.-C. Amann, J. J. Finley, G. Abstreiter & G. Koblmüller, *Nano Lett.* **13**, 6070 (2013).

[23] H. A. Nilsson, P. Caroff, C. Thelander, M. Larsson, J. B. Wagner, L.-E. Wernersson, L. Samuelson & H. Q. Xu, *Nano Lett.* **9**, 3151(2009).

[24] H. A. Nilsson, O. Karlström, M. Larsson, P. Caroff, J. N. Pedersen, L. Samuelson, A. Wacker, L.-E. Wernersson & H. Q. Xu, *Phys. Rev. Lett.* **104**, 186804 (2010).

[25] H. A. Nilsson, P. Samuelsson, P. Caroff & H. Q. Xu, *Nano Lett.* **12**, 228 (2012).

[26] M. T. Deng, C. L. Yu, G. Y. Huang, M. Larsson, P. Caroff & H. Q. Xu, *Nano Lett.* **12**, 6414 (2012).

[27] J.-Y. Wang, S. Huang, G.-Y. Huang, D. Pan, J. Zhao & H. Q. Xu, *Nano Lett.* **17**, 4158 (2017).

[28] S. Li, N. Kang, P. Caroff & H. Q. Xu, *Phys. Rev. B* **95**, 014515 (2017).

[29] M. Keplinger, T. Mårtensson, J. Stangl, E. Wintersberger, B. Mandl, D. Kriegner, V. Holý, G. Bauer, K. Deppert & L. Samuelson, *Nano Lett.* **9**, 5 (2009).

[30] M. Montazeri, M. Fickenscher, L. M. Smith, H. E. Lackson, J. Yarrison-Rice, J. H. Kang, Q. Gao, H. H. Tan, C. Jagadish, Y. Guo, J. Zou, M.-E. Pistol & C. E. Pryor, *Nano Lett.* **10**, 880 (2010).

[31] A. Biermanns, T. Rieger, G. Bussone, U. Pietsch, D. Grützmacher & M. I. Lepsa, *Appl. Phys. Lett.* **102**, 043109 (2013).

[32] L. Dupré, D. Buttard, P. Gentile, Q. B. à la Guillaume, T. Gorisse & H. Renevier, *Phys. Status, Solidi RRL* **8**, 317 (2014).

[33] P. Liu, H. Huang, X. Liu, M. Bai, D. Zhao, Z. Tang, X. Huang, J.-Y. Kim & J. Guo, *Phys. Status Solidi A* **212**, 617 (2015).

[34] P. K. Kasanaboina, S. K. Ojha, S. U. Sami, C. L. Reynolds Jr, Y. Liu & S. Iyer, *Semicond. Sci. Technol.* **30**, 105036 (2015).

[35] D. C. Dillen, F. Wen, K. Kim & E. Tutuc, *Nano Lett.* **16**, 392 (2016).

[36] F. Boxberg, N. Søndergaard & H. Q. Xu, *Nano Lett.* **10**, 1108–1112 (2010).

[37] E. J. Jones, S. Ermez, S. Gradečak, *Nano Letters*, **15**, 7873−7879 (2015).

[38] M. C. Newton, S. J. Leake, R. Harder, I. K. Robinson, *Nature Materials* **9**, 120–124 (2010).

[39] S. A. Dayeh, W. Tang, F. Boioli, K. L. Kavanagh, H. Zheng, J. Wang, N. H. Mack, G. Swadener, J. Y. Huang, L. Miglio, K.-N. Tu, and S. T. Picraux, *Nano Lett.* **13** (5), 1869-1876 (2013).

[40] K. L. Kavanagh, I. Saveliev, M. Blumin, G. Swadener, H. E. Ruda, *J. Appl. Phys.* **111**, 044301 (2012).

[41] F. Boxberg, N. Søndergaard & H. Q. Xu, Adv. Mater. **24**, 4692–4706 (2012).

[42] J. Grönqvist, N. Søndergaard, F. Boxberg, T. Guhr, S. Åberg & H. Q. Xu, *J. Appl. Phys* **106**, 053508 (2009).

[43] M.-E. Pistol, C. Pryor, *Phys. Rev. B* **78**, 115319 (2008).

[44] M. H. Magnusson, K. Deppert, J.-O. Malm, J.-O. Bovin & L. Samuelson, *Nanostructured Materials* **12**, 45-48 (1999).

[45] J. Wallentin, K. Mergenthaler, M. Ek, L. R. Wallenberg, L. Samuelson, K. Deppert, M.-E. Pistol & M. T. Borgström, *Nano Lett.* **11**, 2286-2290 (2011).

[46] J. Wallentin, M. E. Messing, E. Trygg, L. Samuelson, K. Deppert & M. T. Borgström, *J. Crystal Growth* **331**, 8–14 (2011).

[47] A. Berg, K. Mergenthaler, M. Ek, M.-E. Pistol, L. R. Wallenberg & M. T. Borgström, *Nanotechnology* **25**, 505601 (2014).

[48] B. M. Borg, K. A. Dick, J. Eymery & L.-E. Wernersson, *Appl. Phys. Lett.* **98**, 113104 (2011).

[49] D. Kriegner, C. Panse, B. Mandl, K. A. Dick, M. Keplinger, J. M. Persson, P. Caroff, D. Ercolani, L. Sorba, F. Bechstedt, J. Stangl & G. Bauer, *Nano Lett.* **11**, 1483–1489 (2011).

[50] D. Kriegner, E. Wintersberger, K. Kawaguchi, J. Wallentin, M. T. Borgström & J. Stangl, Nanotechnology **22**, 425704 (2011).

[51] S. Q. Wang & H. Q. Ye, *Phys. Status Solidi (b)* **240**, 45 (2003).

[52] C. Hajlaoui, L. Pedesseau, F. Raouafi, F. Ben Cheikh Larbi, J. Even & J.-M. Jancu, *J. Phys. D: Appl. Phys.* **46**, 505106 (2013).

# Supporting Information for

# Measurements of strain and bandgap of coherently epitaxially grown wurtzite InAsP-InP core-shell nanowires

D. J .O. Göransson[1], M. T. Borgström[1], Y. Q. Huang[2], M. E Messing[1], D. Hessman[1], I. A. Buyanova[2], W. M. Chen[2], H. Q. Xu[1,3,4,*]

[1]*NanoLund and Division of Solid State Physics, Lund University, Box 118, S-22100 Lund, Sweden*

[2]*Department of Physics, Chemistry and Biology, Linköping University, S-581 83 Linköping, Sweden*

[3] *Beijing Key Laboratory of Quantum Devices, Key Laboratory for the Physics and Chemistry of Nanodevices, and Department of Electronics, Peking University, Beijing 100871, China*

[4] *Beijing Academy of Quantum Information Sciences, West Bld. #3, No.10 Xibeiwang East Rd., Haidian District, Beijing 100193, China*

[*] *Email address: hqxu@pku.edu.cn and hongqi.xu@ftf.lth.se*

**This supporting information contains the following four sections:**

Section S1. Photoluminescence measurements of single InAsP-InP core-shell nanowires

Section S2. Photoluminescence measurements of single reference InAsP nanowires

Section S3. µ-Raman measurements of single InAsP-InP core-shell and reference InAsP nanowires

Section S4. XRD spectra of two reference samples measured before and after removal of the InAsP nanowires

## Section S1. Photoluminescence measurements of single InAsP-InP core-shell nanowires

The micro-photoluminescence (µPL) spectra of individual nanowires were measured with varying excitation laser power. The PL peak energies were then extracted by fitting a Gaussian curve to the shape of each peak. The excitation power was reduced until the peak positions were found to be stable for each nanowire. The spectra from InAsP-InP core-shell nanowires with shell growth times of 15, 22, 30, 45, 60 and 90 s are shown in figures S1-1, S1-3, S1-5, S1-7, S1-9 and S1-11, respectively. The power dependences of the extracted peak positions of these nanowires are shown in Figures S1-2, S1-4, S1-6, S1-8, S1-10 and S1-12. Table S1 shows the average values of strain, diameter and µPL peak energy of the individual InAsP-InP core-shell nanowires studied in the main article together with reference InAsP nanowires without growth of shells.

Table S1-1. Average strain extracted from X-ray diffraction (XRD) data, average diameter determined using scanning electron microscope (SEM), and average PL peak energy found at the lowest excitation power for individual InAsP-InP core-shell nanowires studied in the main article together with reference InAsP nanowires without shells.

| InP shell growth time (s) | 0 | 15 | 22 | 30 | 45 | 60 | 90 |
|---|---|---|---|---|---|---|---|
| Strain in the InAsP core ($\varepsilon_{zz}$) | 0 | -0.00115 | -0.00355 | -0.00429 | -0.00547 | -0.00657 | -0.00845 |
| Average diameter (nm) | 42.5 | 50.2 | 54.0 | 53.3 | 65.3 | 79.3 | 99.3 |
| Average PL energy at the lowest excitation power (eV) | 1.086 | 1.092 | 1.075 | 1.124 | 1.122 | 1.149 | 1.159 |

Figure S1-1 shows the µPL measuremets of 5 individual InAsP-InP core-shell nanowires with a shell growth time of 15 s at different laser excitation power intensities. Figure S1-2 shows the PL peak energy extracted from the measured spectra as a function of the excitation power intensity.

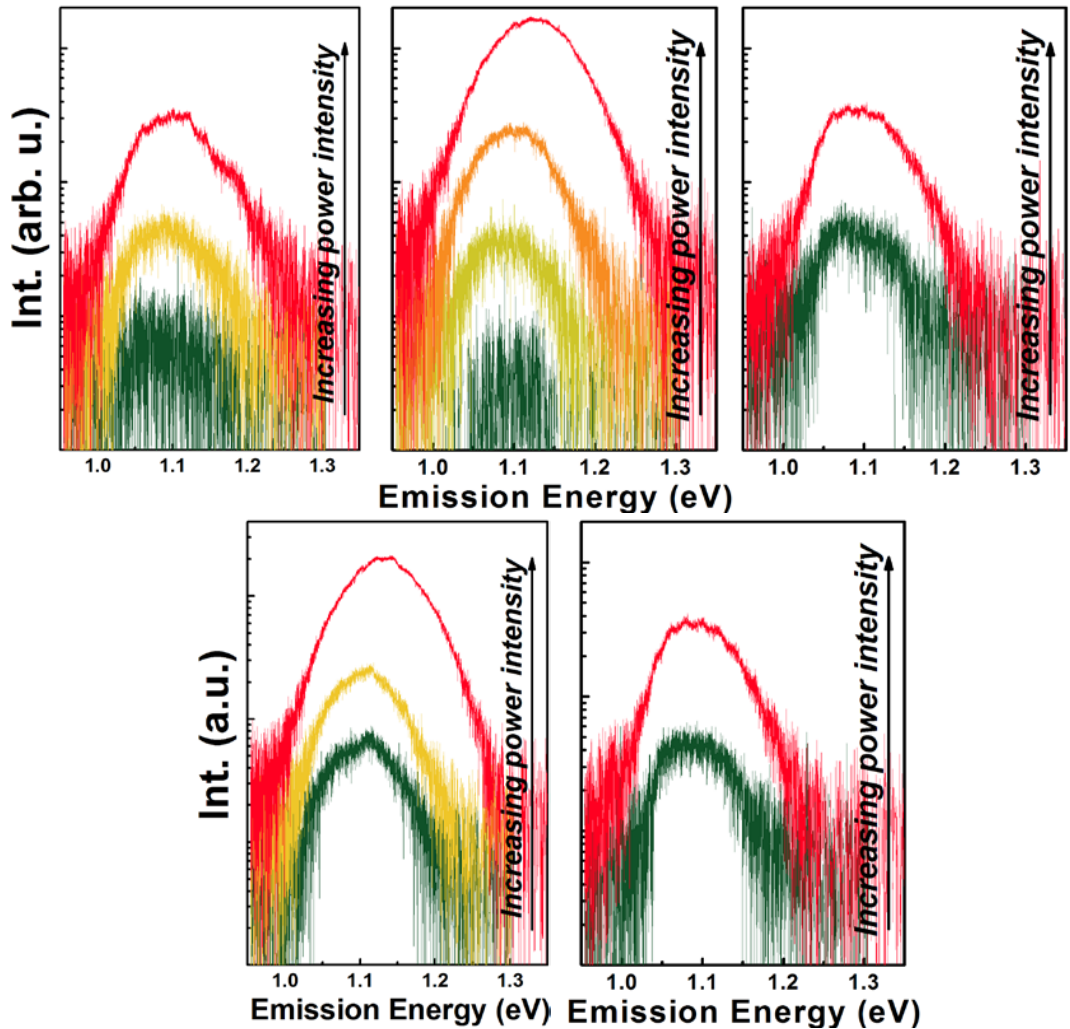

Figure S1-1. µPL spectra measured at 4 K for 5 individual InAsP-InP core-shell nanowires with a shell growth time of 15 s. For each nanowire, PL spectra were measured with the excitation laser wavelength fixed at 659 nm but with different excitation power intensities.

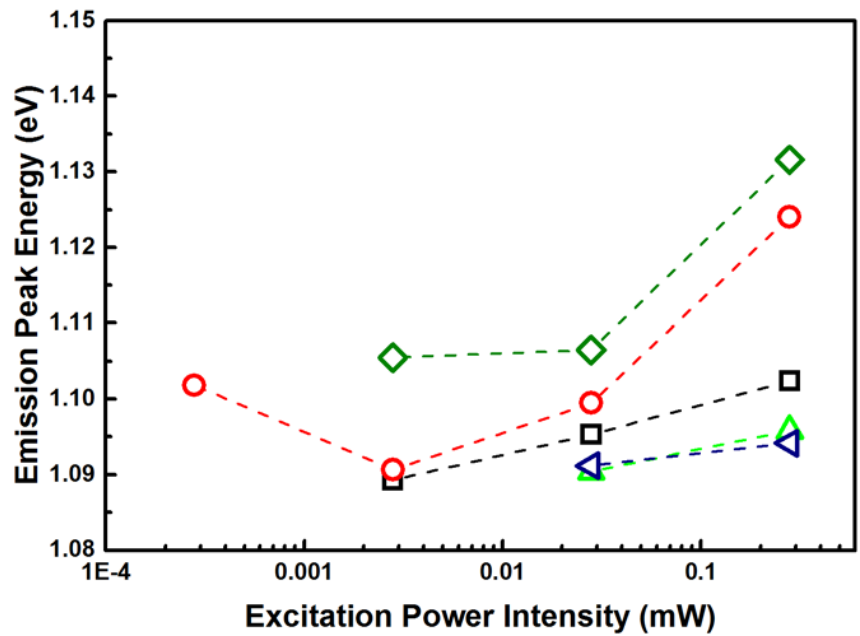

Figure S1-2. Peak emission energy of the InAsP-InP core-shell nanowires with a shell growth time of 15 s, extracted from the spectra shown in Figure S1-1, as a function of the laser excitation power intensity. Each color corresponds to the measurements for a single nanowire.

Figure S1-3 shows the µPL measuremets of 6 individual InAsP-InP core-shell nanowires with a shell growth time of 22 s at different laser excitation power intensities. Figure S1-4 shows the PL peak energy extracted from the measured spectra as a function of the excitation power intensity.

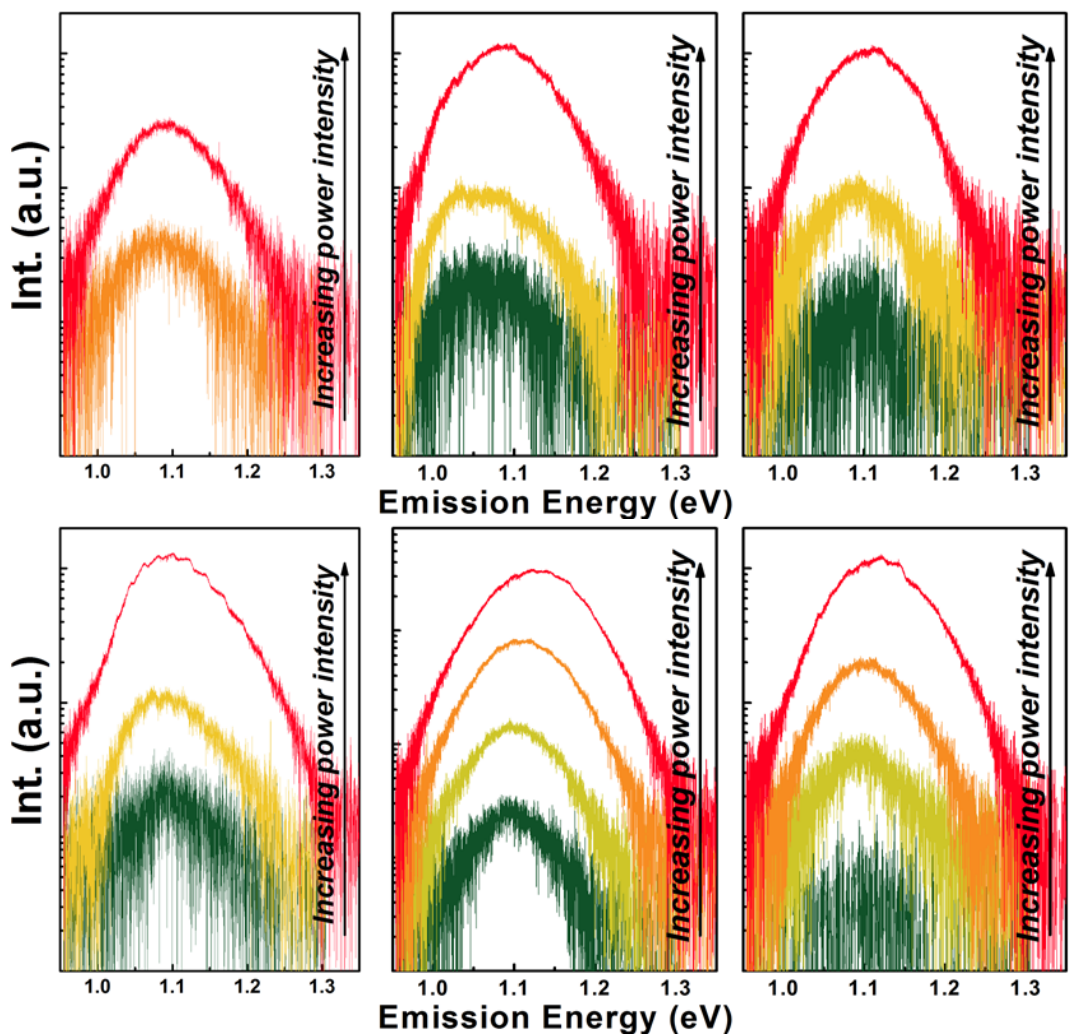

Figure S1-3. µPL spectra measured at 4 K for 6 individual InAsP-InP core-shell nanowires with a shell growth time of 22 s. For each nanowire, PL spectra were measured with the excitation laser wavelength fixed at 659 nm but with different excitation power intensities.

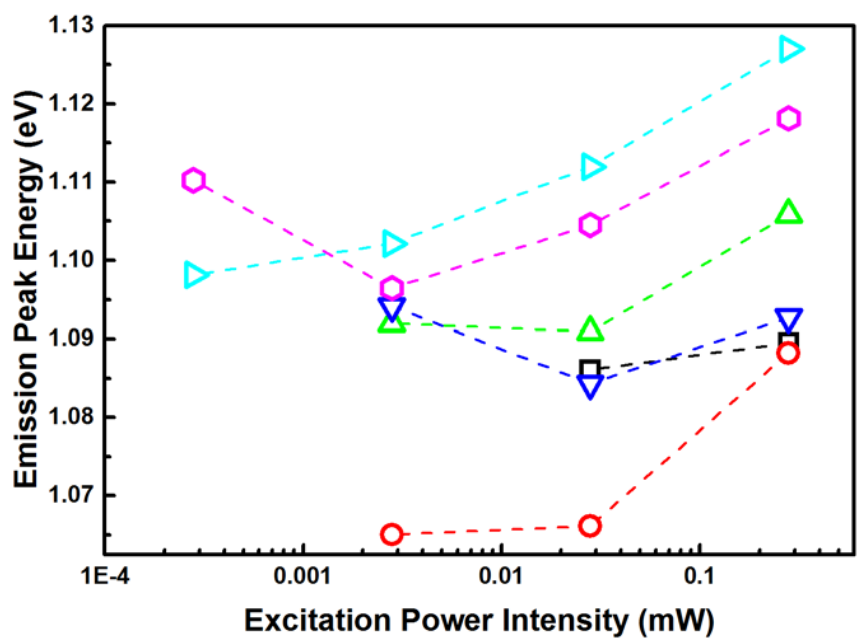

Figure S1-4. Peak emission energy of the InAsP-InP core-shell nanowires with a shell growth time of 22 s, extracted from the spectra shown in Figure S1-3, as a function of the laser excitation power intensity. Each color corresponds to the measurements for a single nanowire.

Figure S1-5 shows the µPL measurements of 6 individual InAsP-InP core-shell nanowires with a shell growth time of 30 s at different laser excitation power intensities. Figure S1-6 shows the PL peak energy extracted from the measured spectra as a function of the excitation power intensity.

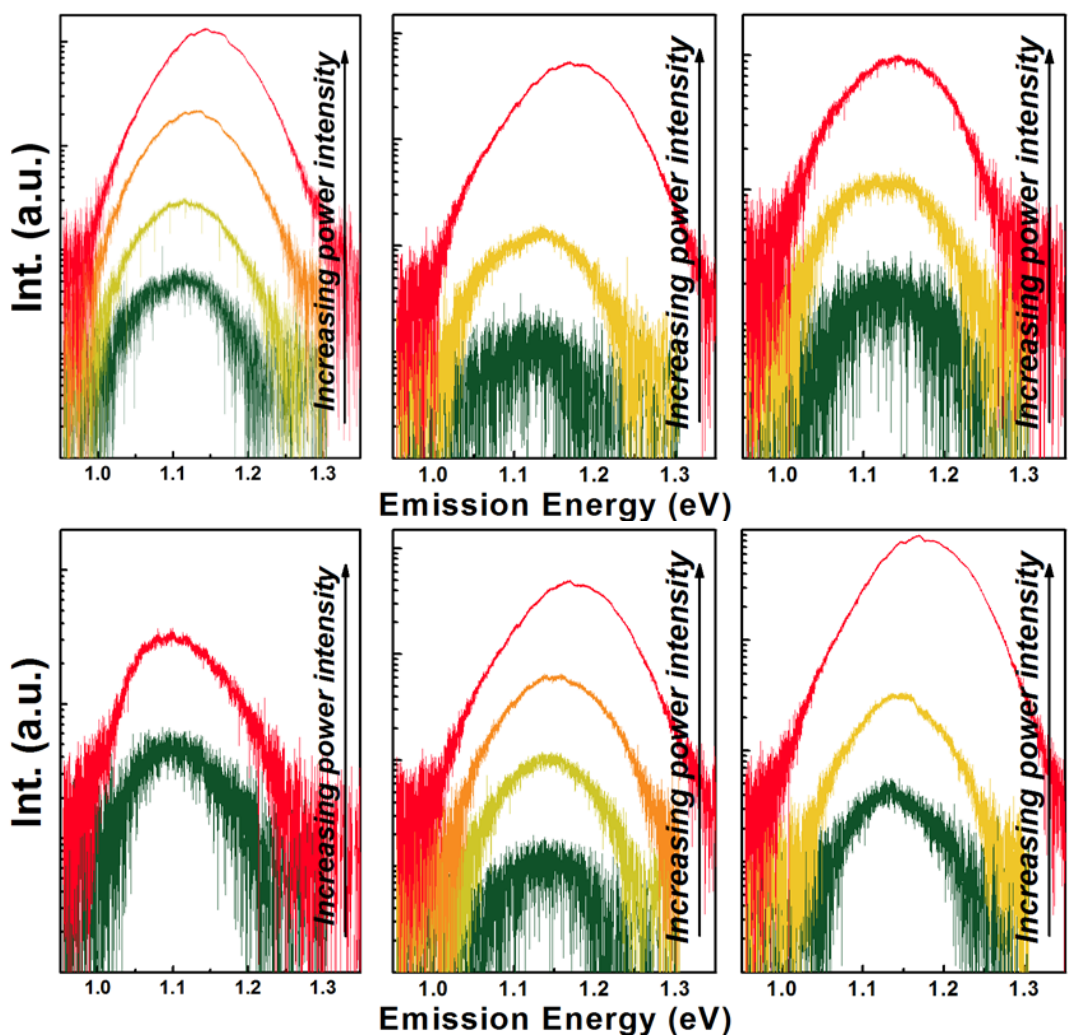

Figure S1-5. µPL spectra measured at 4 K for 6 individual InAsP-InP core-shell nanowires with a shell growth time of 30 s. For each nanowire, PL spectra were measured with the excitation laser wavelength fixed at 659 nm but with different excitation power intensities.

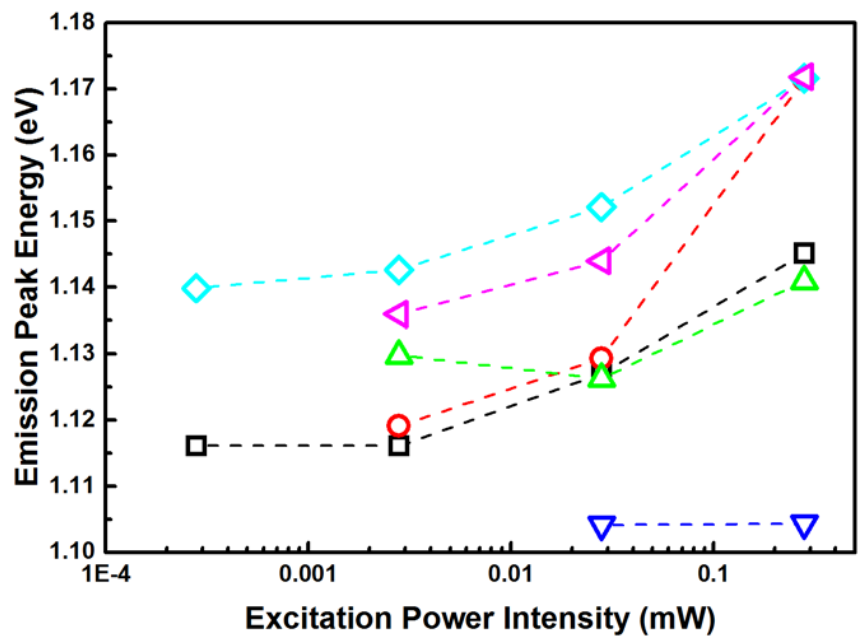

Figure S1-6. Peak emission energy of the InAsP-InP core-shell nanowires with a shell growth time of 30 s, extracted from the spectra shown in Figure S1-5, as a function of the laser excitation power intensity. Each color corresponds to the measurements for a single nanowire.

Figure S1-7 displays the μPL measurements of 5 individual InAsP-InP core-shell nanowires with a shell growth time of 45 s at different laser excitation power intensities. In the spectra obtained at high excitation powers, two PL bands are visible. However, in the spectra obtained at the lowest excitation power, only the low energy band is luminescent due to carrier relaxation to this band. Figure S1-8 shows the peak position of the high energy PL band extracted from the measured spectra in Figure S1-7 as a function of the excitation power intensity. The peak position of the high energy band is extracted by subtracting the low energy band component from the total spectra. The low energy band component at a given excitation power is estimated by taking the corresponding low energy band at the lowest excitation power and then scaling it up with the excitation power to the given one.

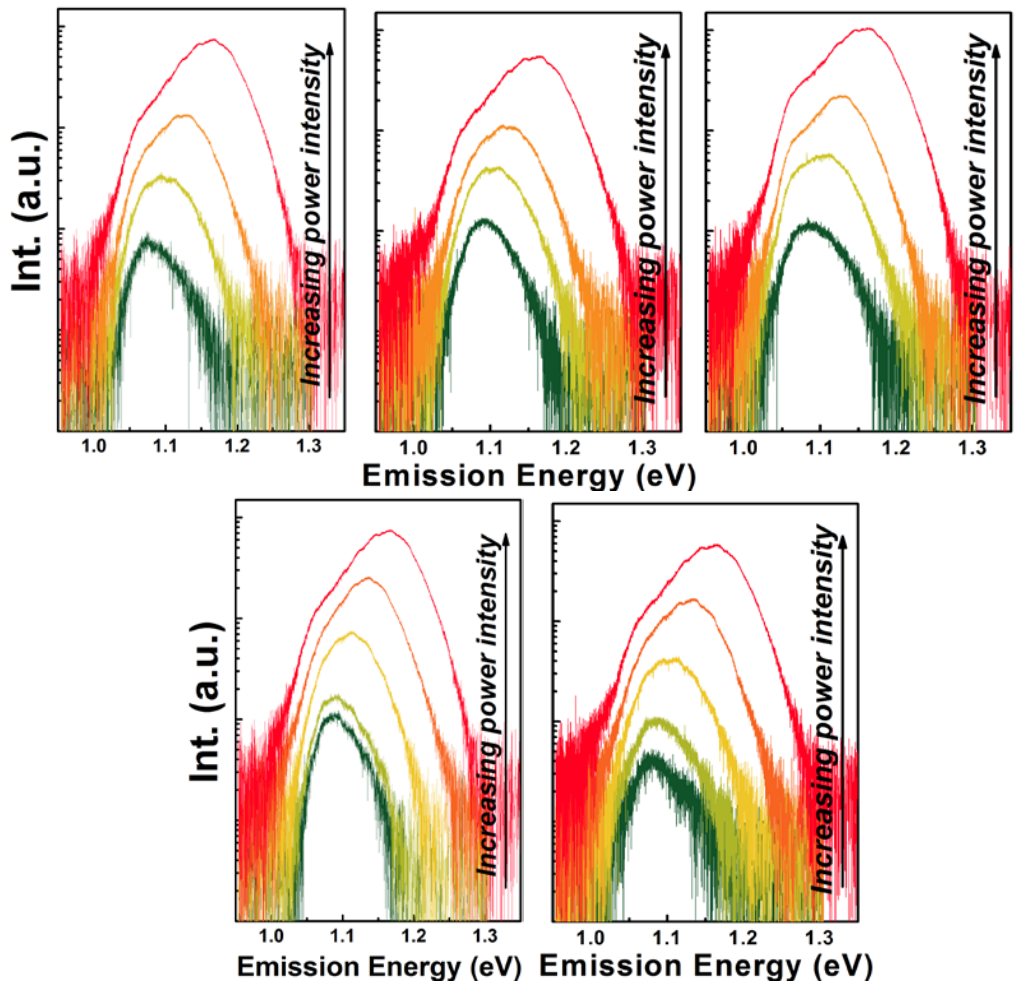

Figure S1-7. µPL spectra measured at 4 K for 5 individual InAsP-InP core-shell nanowires with a shell growth time of 45 s. For each nanowire, PL spectra were measured with the excitation laser wavelength fixed at 659 nm but with different excitation power intensities.

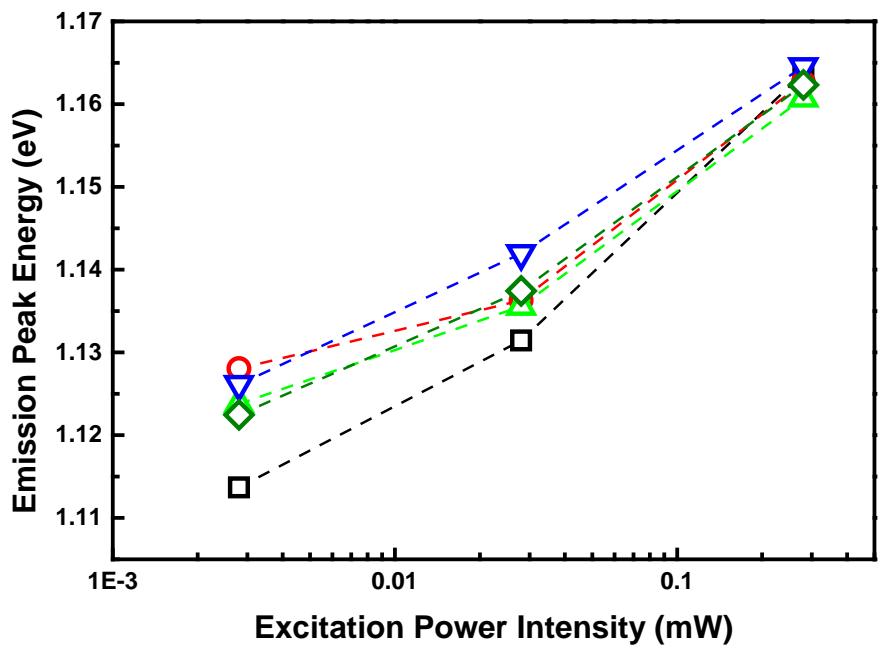

Figure S1-8. Peak emission energy of the InAsP-InP core-shell nanowires with a shell growth time of 45 s, extracted from the high energy band in the spectra shown in Figure S1-7, as a function of the laser excitation power intensity. Each color corresponds to the measurements for a single nanowire.

Figure S1-9 shows the µPL measuremets of 5 individual InAsP-InP core-shell nanowires with a shell growth time of 60 s at different laser excitation power intensities. Figure S1-10 shows the PL peak energy extracted from the measured spectra as a function of the excitation power intensity.

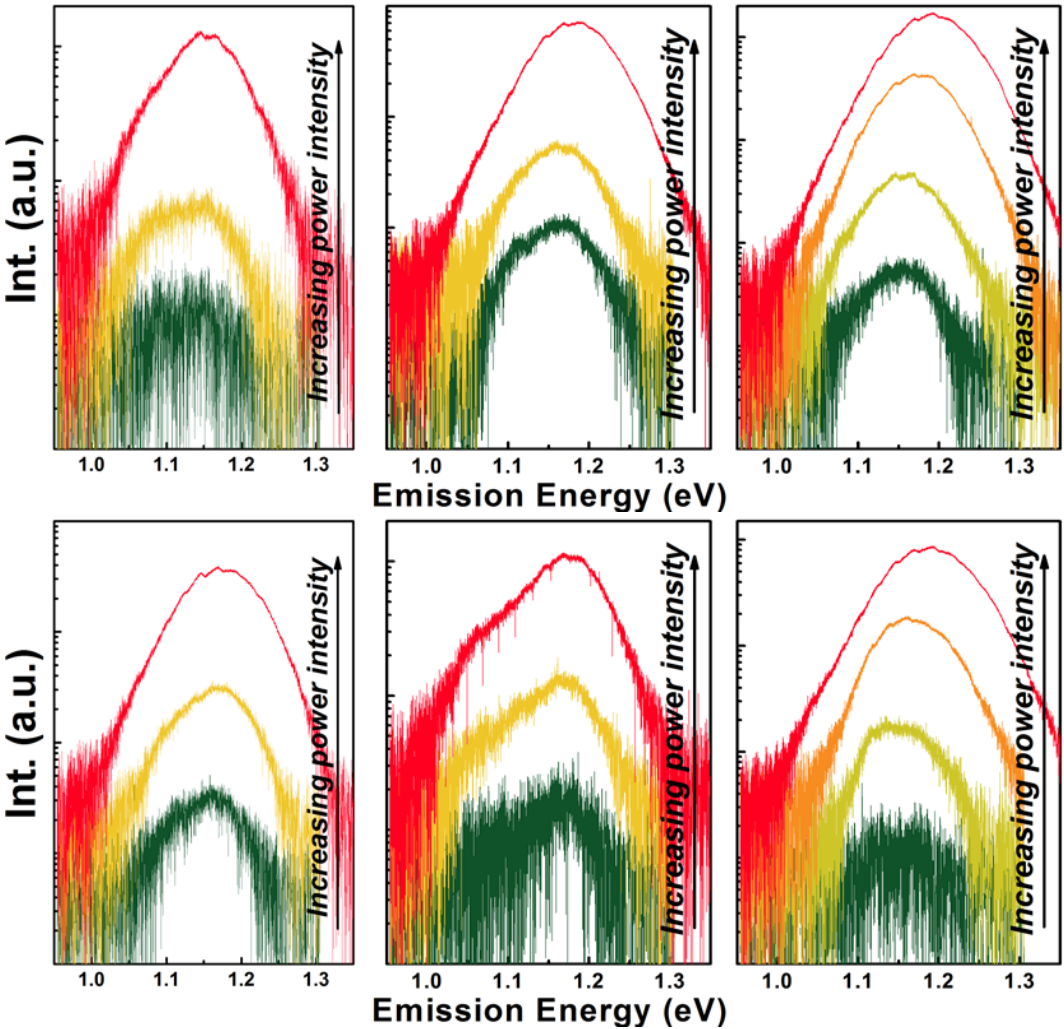

Figure S1-9. µPL spectra measured at 4 K for 6 individual InAsP-InP core-shell nanowires with a shell growth time of 60 s. For each nanowire, PL spectra were measured with the excitation laser wavelength fixed at 659 nm but with different excitation power intensities.

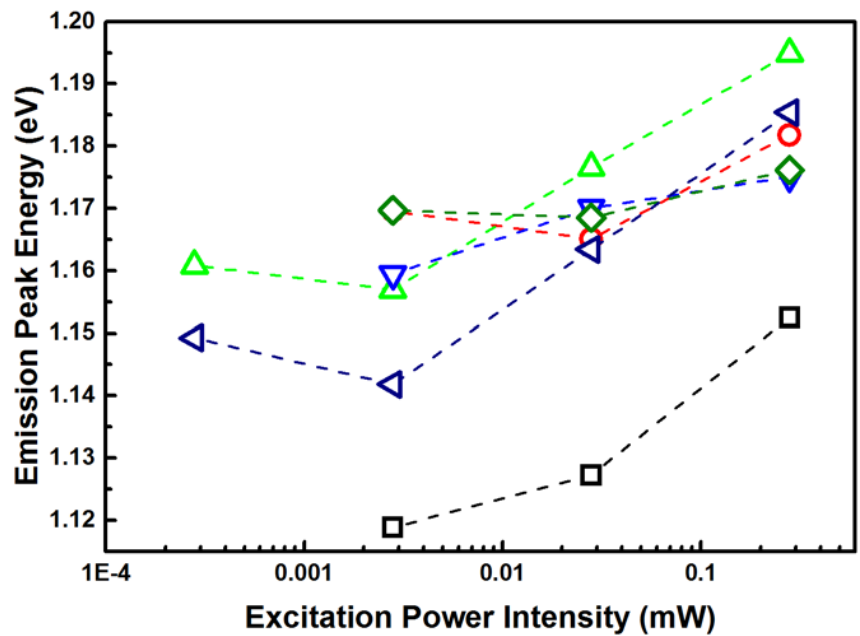

Figure S1-10. Peak emission energy of the InAsP-InP core-shell nanowires with a shell growth time of 60 s, extracted from the spectra shown in Figure S1-9, as a function of the laser excitation power intensity. Each color corresponds to the measurements for a single nanowire.

Figure S1-11 shows the µPL measuremets of 5 individual InAsP-InP core-shell nanowires with a shell growth time of 90 s at different laser excitation power intensities. Figure S1-12 shows the PL peak energy extracted from the measured spectra as a function of the excitation power intensity.

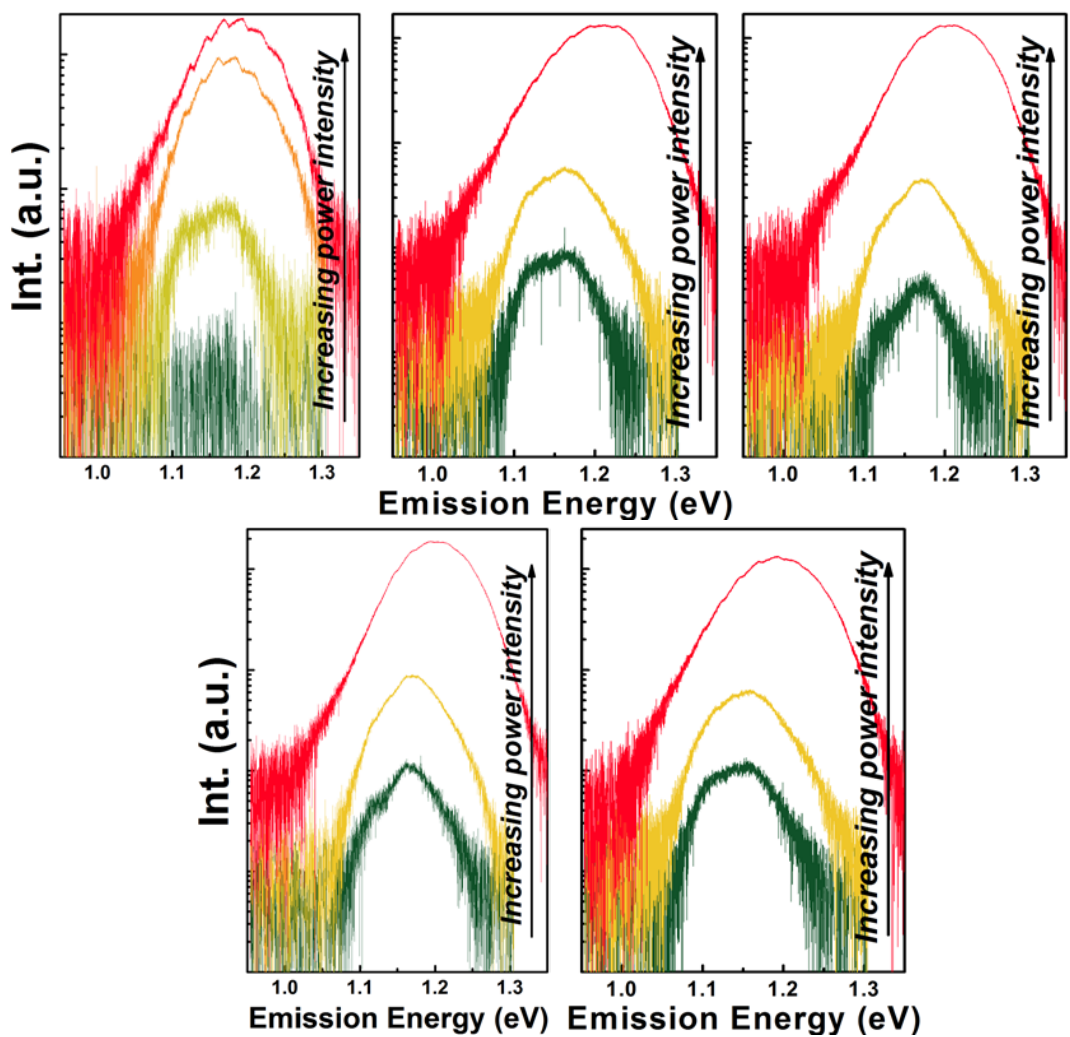

Figure S1-11. µPL spectra measured at 4 K for 5 individual InAsP-InP core-shell nanowires with a shell growth time of 90 s. For each nanowire, PL spectra were measured with the excitation laser wavelength fixed at 659 nm but with different excitation power intensities.

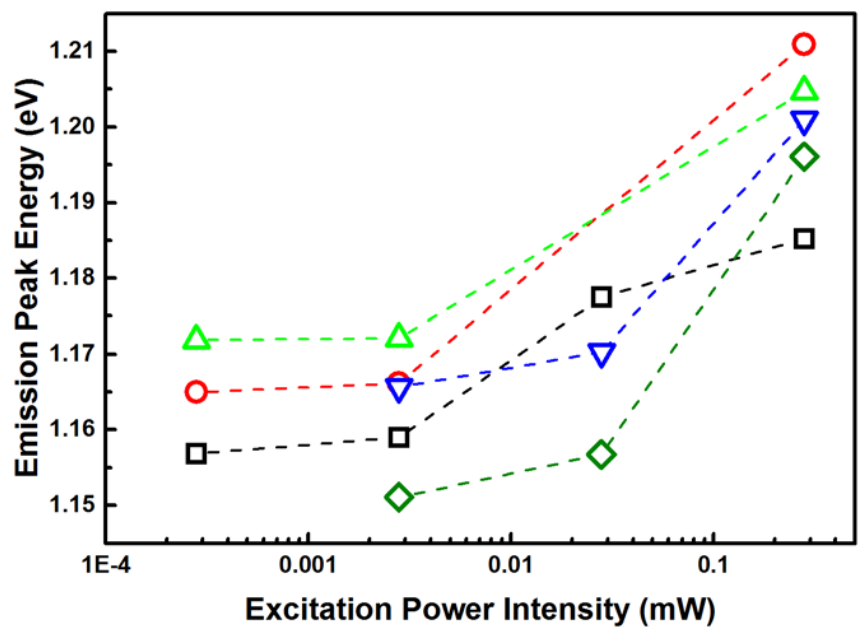

Figure S1-12. Peak emission energy of the InAsP-InP core-shell nanowires with a shell growth time of 90 s, extracted from the spectra shown in Figure S1-11, as a function of the laser excitation power intensity. Each color corresponds to the measurements for a single nanowire.

**Section S2. Photoluminescence measurements of single reference InAsP nanowires**

Four reference samples of InAsP nanowires were grown without InP shells. In order to check and calibrate the growth condition for the InAsP cores in different InAsP-InP core-shell nanowire samples, these reference samplers were grown between two growth runs of InAsP-InP core-shell nanowires . In Figures S2-1, S2-2, S2-3 and S2-4, the µPL spectra measured for eight single InAsP nanowires (with two from one reference sample) at 4 K are shown. In Figure S2-5, the PL energies extracted from all measured reference nanowires are shown.

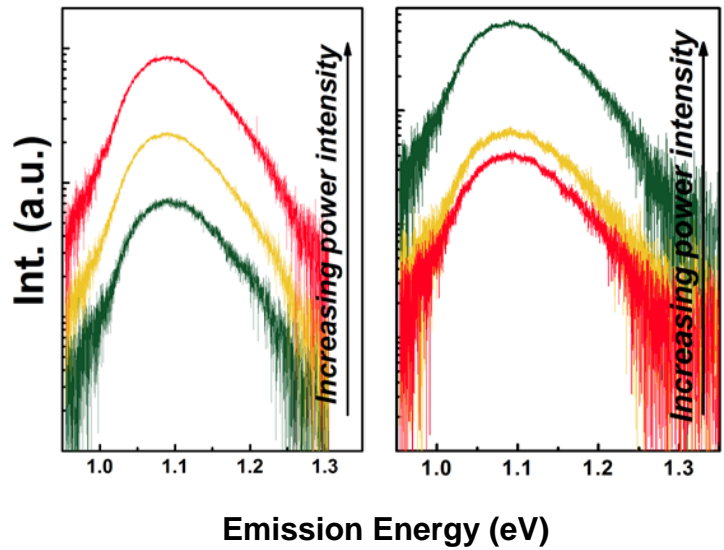

Figure S2-1. µPL spectra measured at 4 K for two individual InAsP nanowires from reference sample 1. For each nanowire, PL spectra were measured with the excitation laser wavelength fixed at 659 nm but with different excitation power intensities.

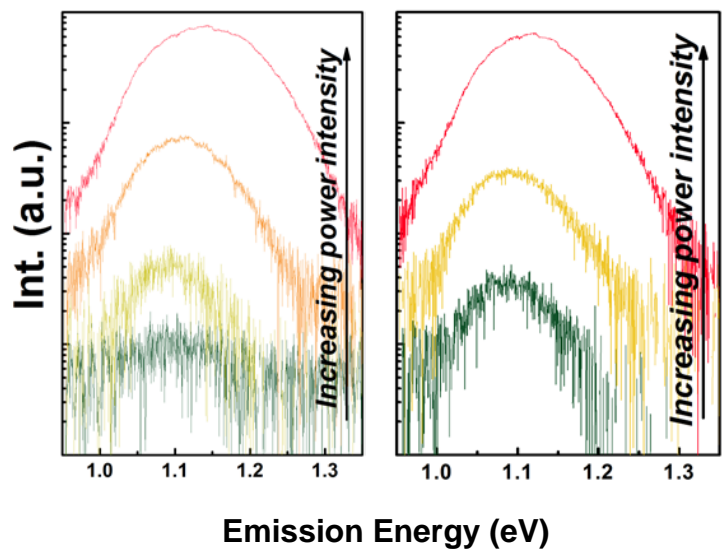

Figure S2-2. µPL spectra measured at 4 K for two individual InAsP nanowires from reference sample 2. For each nanowire, PL spectra were measured with the excitation laser wavelength fixed at 659 nm but with different excitation power intensities.

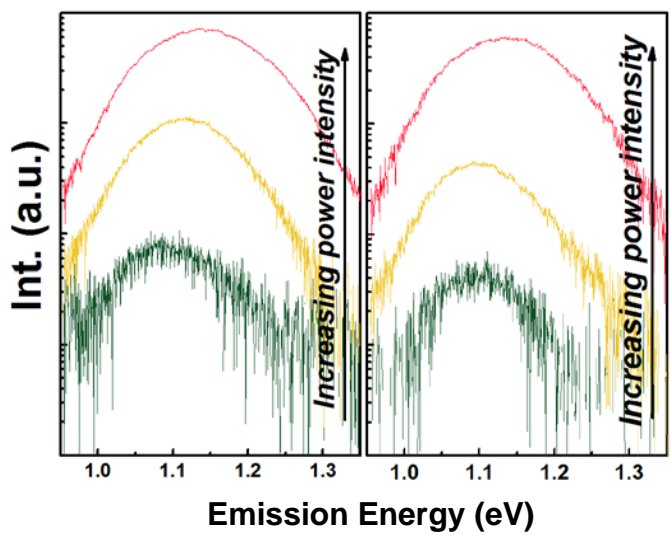

Figure S2-3. µPL spectra measured at 4 K for two individual InAsP nanowires from reference sample 3. For each nanowire, PL spectra were measured with the excitation laser wavelength fixed at 659 nm but with different excitation power intensities.

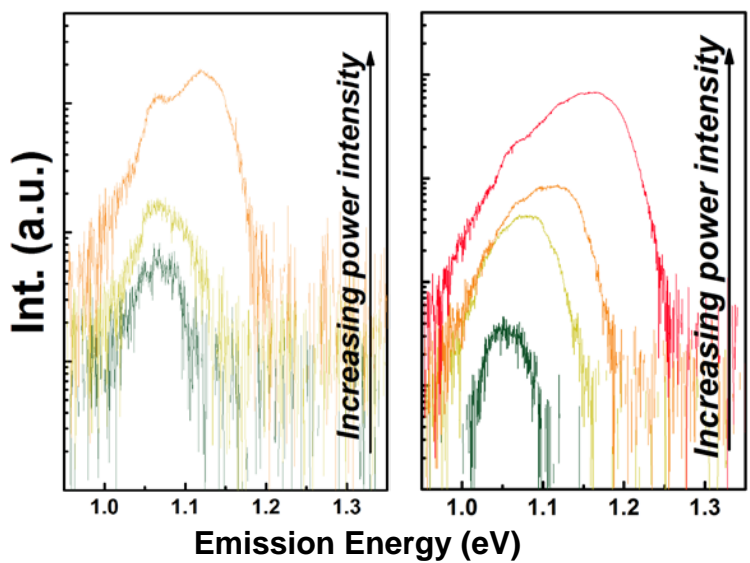

Figure S2-4. µPL spectra measured at 4 K for two individual InAsP nanowires from reference sample 4. For each nanowire, PL spectra were measured with the excitation laser wavelength fixed at 659 nm but with different excitation power intensities.

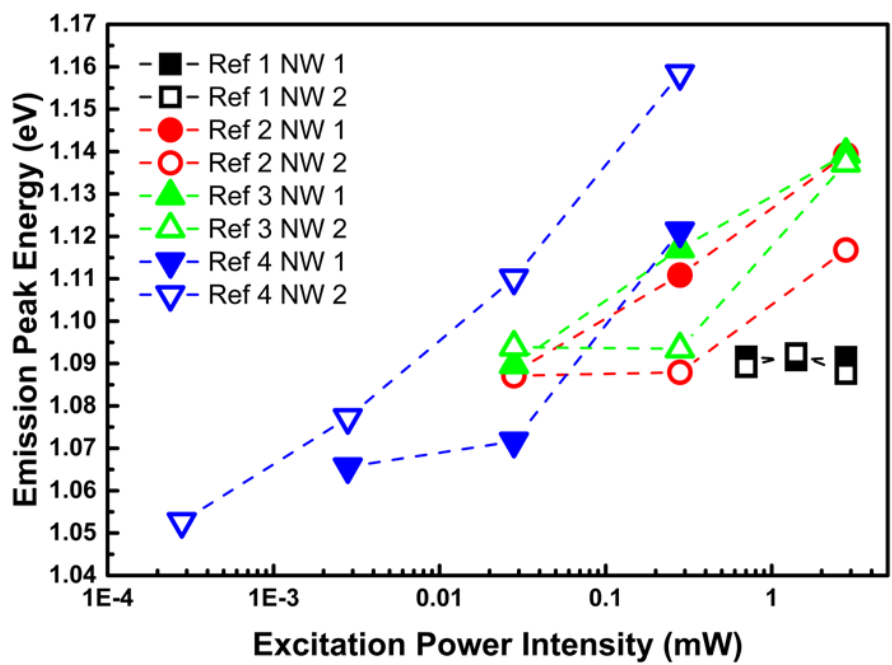

Figure S2-5. Peak emission energy of the eight InAsP nanowires from the four reference samples, extracted from the spectra shown in Figures S2-1 to S2-4, as a function of the laser excitation power intensity.

## Section S3. μ-Raman measurements of single InAsP-InP core-shell and reference InAsP nanowires

The μ-Raman spectroscopic measurements were carried out at room temperature in a confocal microscope. The nanowires were mechanically transferred to glass substrates and isolated individual nanowires were then located by optical microscopic imaging. The glass substrates were chosen here to ensure a clean background in the interested spectral range of 260 to 380 $cm^{-1}$. A solid-state laser beam with emitting wavelength of 659 nm was guided and focused onto the nanowires through a 100 magnification objective lens and the Raman signals were collected in a backscattering geometry through the same objective lens. The laser spot diameter was about 0.9 μm and the excitation power density was about $4 \times 10^4 \, W \cdot cm^{-2}$. The Raman signals were then collected to pass through an edge filter with a proper cut-off wavelength, which filtered out the elastic-scattered light and ensured a reliable Raman characterization for wavenumbers above 100 $cm^{-1}$. The Raman signals were dispersed with a 0.8-m monochromator and detected by a Si CCD camera. We note that the measurements were done without distinguishing the polarization of the phonon modes.

In Figure S3-1 (a), we have plotted out the representative μ-Raman spectra obtained from single InAsP-InP core-shell nanowire with different shell growth times. The spectrum of a reference InAsP nanowire without shell growth is shown in the figure. The spectra are shifted vertically for easy comparison. Starting with the reference nanowire, two peaks are clearly visible, whose peak positions are marked with dashed lines at 300.2 and 330.6 $cm^{-1}$, respectively. Because of the absence of InP shell, the observed Raman modes are expected to be contributed solely by the strain-free InAsP nanowire. We find that the Raman shifts match nicely with the zone center longitudinal optical (LO) phonon mode and transverse optical (TO) phonon mode of $InAs_xP_{1-x}$ with As composition x=0.34 [1]. This result confirms the As concentration determined by our XRD measurements. With increasing shell growth time, the LO phonon peak shifts appreciably towards higher energy, whereas the TO peak shows slightly shift towards the lower energy. For the core-shell nanowire grown for 30 s and thus of 53 nm in total nanowire diameter, the TO and LO peaks are find at 335.7 and 298.1 $cm^{-1}$, respectively. For the nanowire of 65 nm in total nanowire diameter obtained by increasing shell growth time to 45 s, the TO and LO peaks became strongly asymmetric such that they can no longer be fitted with a single Lorentzian. In Figure S3-1(b), we show the spectrum of the nanowire in a vertically enlarged scale. Here, at least four phonon modes, which we label as TO2, TO1, LO1 and LO2, can be identified. The corresponding Raman shifts, which are determined from fitting by Lorentzians, are 296.7, 302.6, 334.9 and 338.6 $cm^{-1}$. As shown in Ref. [2], the deconvolution of the TO peak is not a trivial task. Even for pure wurtzite InAsP, the TO peak may consist of several Raman modes, including an A1(TO) and an E1(TO) mode with roughly same Raman shift and an E2h(TO) mode which is about 4 $cm^{-1}$ separated from the A1(TO) and the E1(TO) mode. In our case, the task is further complicated by the contribution from the wurtzite InP shell as well as strains at the core-shell interface. Fortunately, the situation is comprehensible for the LO peak, since the LO peak is expected to arise only from the singly degenerated A1(LO) mode. The appearance of the double LO peak in Figure S3-1(b) can consequently be assigned to the LO phonon in the shell and the core region of the nanowire, taking into account that the strain is uniform in the core of the nanowire. With further increasing shell thickness, the symmetry of the Raman peaks is restored and only a single LO and a single TO component can be identified in the measured Raman spectrum. For the nanowires with a shell growth time exceeding 45 s, the observed LO and TO peaks are found to appear at wavenumbers around 336 $cm^{-1}$ and 298.5 $cm^{-1}$, and become insensitive to the increase of shell thickness.

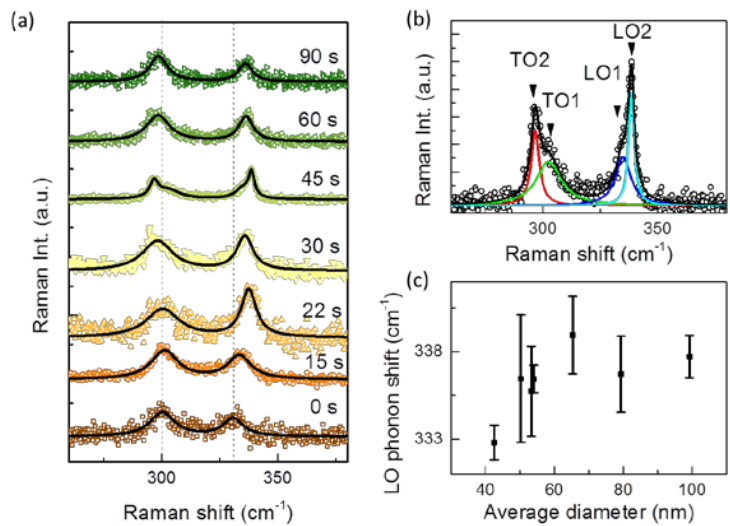

Figure S3-1. (a) Representative Raman spectra (symbols) of the core-shell nanowire samples with different shell growth times. The spectra are fitted with Lorentzians, which correspond to the contributions from the LO and TO phonons of InAsP and/or InP, (black curves). The shell growth times are in the order of 0, 15, 22, 30, 45, 60, 90 s from the bottom to the top and the corresponding average diameters are in the order of 42.5, 50.2, 54.0, 53.3, 65.3, 79.9, and 99.3 nm. For sample with the 45 s shell growth time, the LO and TO phonon peaks split into two components each. (b) Detailed Raman spectra (symbols)and fitting results (black curve) of the sample with the 45 s shell growth time. Here each peak is fitted with two Lorentzions. (c) Extracted LO phonon Raman shift as a function of the average diameter of the core-shell nanowires. The mean value and standard deviation of the Raman shifts are obtained from statistics of 7 nanowires for each data point. The averaged nanowire diameter for each sample is obtained from the SEM measurements of approximately 20 nanowires on the sample.

To understand the shell thickness dependence of the Raman spectra, one needs to consider the competition between core and shell region of the nanowire. For a InAsP-InP core-shell nanowire with a small shell thickness, the Raman spectrum is dominated by the phonon modes of the InAsP core. Prolonging the shell growth time lead to built-up of uniaxial strain along the nanowire growth direction. The negative strain of nanowire core induces the blue shift of the LO phonon mode, which is exactly what we have seen in the samples with shell growth times less than 45 s. For the nanowire with the shell grown for 45 s, the Raman signals from the shell region becomes comparable in intensity to that from the core region, which lead to the appearance of the double LO peak as highlighted in Figure S3-1(b). With a further increase of the shell thickness, the Raman spectra originate mainly from the shell region, as the excitation is largely attenuated and the scattered light is difficult to extract out from the nanowire core. This explains the shell thickness independence of the Raman peak positions. To show this as a general trend, we have thus carried out statistical analysis of the Raman results from the measurements of seven nanowire samples with different averaged diameters. For each sample, seven nanowires were studied. In Figure S3-1(c), we have plotted out the average LO phonon peak position and the corresponding standard deviation as a function of the averaged nanowire diameter. It is evident that the overall LO phonon shift follows the aforementioned trend that the LO phonon peak position shows a continuous blue shift for nanowires with a small shell thickness and stays approximately at the same wavenumber for nanowires with a thicker shell.

References

[1] Carles, R.; Saint-Cricq, N.; Renucci, J. B.; Nicholas, R. J., J. Phys. C: Solid State Phys. 13 (5), 899–910 (1980).

[2] Gadret, E.G.; Jr, M. M. de Lima; Madureira, J. R.; Chiaramonte, T.; Cotta, M. A.; Iikawa, F.; Cantarero, A., Appl. Phys. Lett. 102 (11), 122101 (2013).

# Section S4. XRD spectra of two reference samples measured before and after removal of the InAsP nanowires

XRD spectra (figure S4-1) were acquired before and after the nanowires were removed from the substrate using a clean-room paper cloth. The peak at 25.85-25.9 degrees is not visible after the removal of the nanowires, supporting the conclusion that the peak originated from the vertically aligned nanowires on the substrates.

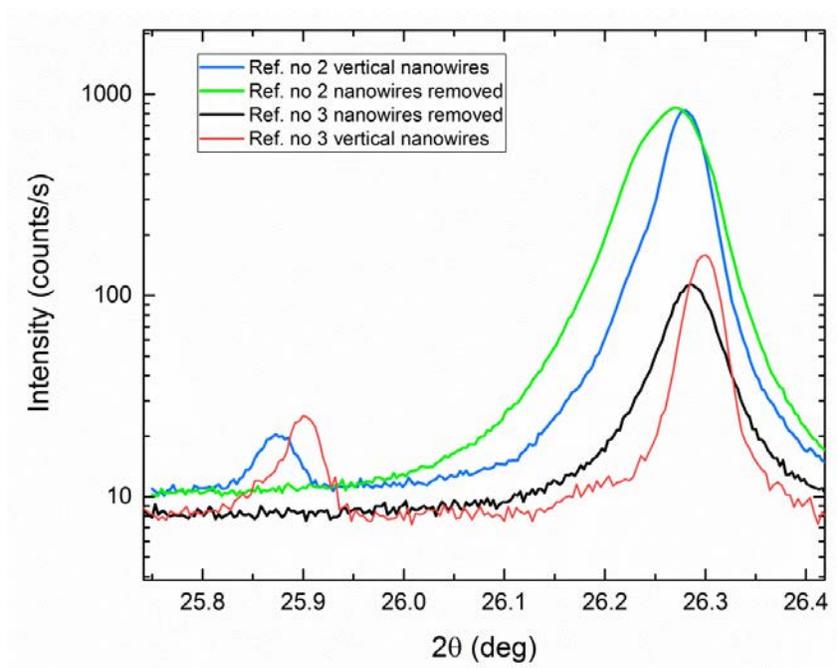

Figure S4-1. XRD spectra acquired in a *2θ-ω* setup with an *ω* offset of -0.1 degrees. The spectra were taken before and after the nanowires were removed from the substrates.